\documentclass[letterpaper, conference]{IEEEtran}
\usepackage{amsmath,amsfonts}
\usepackage{algorithmic}
\usepackage{algorithm}
\usepackage{array}
\usepackage{textcomp}
\usepackage{stfloats}
\usepackage{url}
\usepackage{verbatim}
\usepackage{graphicx}
\usepackage{cite}
\usepackage{booktabs}
\usepackage{xcolor}
\usepackage{multirow}
\usepackage{setspace}
\usepackage[letterpaper, right=0.64in, left=0.64in, top=1.05in, bottom=1.05in]{geometry}  

\usepackage{lettrine}
\begin{document}

\title{Edge Agentic AI Framework for Autonomous Network Optimisation in O-RAN} 
\author{
  Abdelaziz Salama\textsuperscript{$\star$} \quad Zeinab Nezami \textsuperscript{$\star$} \quad Mohammed M. H. Qazzaz\textsuperscript{$\star$}  \quad Maryam Hafeez\textsuperscript{$\star$}
  \quad Syed Ali Raza Zaidi\textsuperscript{$\star$}\\ 
  \textsuperscript{$\star$}School of Electrical and Electronic Engineering, University of Leeds, Leeds, UK\\
  \textsuperscript{$\star$}Corresponding author: Abdelaziz Salama (A.M.Salama@Leeds.ac.uk)
\thanks{This research was funded by EPSRC CHEDDAR (EP/X040518/1), UKRI Grant EP/X039161/1, and MSCA Horizon EU Grant 101086218.}

}

\maketitle

\begin{abstract}
The deployment of AI agents within legacy Radio Access Network (RAN) infrastructure poses significant safety and reliability challenges for future 6G networks. This paper presents a novel Edge AI framework for autonomous network optimisation in Open RAN environments, addressing these challenges through three core innovations: (1) a persona-based multi-tools architecture enabling distributed, context-aware decision-making; (2) proactive anomaly detection agent powered by traffic predictive tool; and (3) a safety, aligned reward mechanism that balances performance with operational stability.

Integrated into the RAN Intelligent Controller (RIC), our framework leverages multimodal data fusion, including network KPIs, a traffic prediction model, and external information sources, to anticipate and respond to dynamic network conditions. Extensive evaluation using realistic 5G scenarios demonstrates that the edge framework achieves zero network outages under high-stress conditions, compared to 8.4\% for traditional fixed-power networks and 3.3\% for large language model (LLM) agent-based approaches, while maintaining near real-time responsiveness and consistent QoS. These results establish that, when equipped with the right tools and contextual awareness, AI agents can be safely and effectively deployed in critical network infrastructure, laying the framework for intelligent and autonomous 5G and beyond network operations.
\end{abstract}


\section{Introduction}

\lettrine{T}{he} transition to 6G wireless networks introduces a fundamental architectural and operational evolution, moving beyond speed and capacity improvements towards intelligent and autonomous service delivery across diverse domains such as extended reality, digital twins, and autonomous vehicles~\cite{raddo2021transition}. In this vision, artificial intelligence, such as large language models (LLMs), becomes a native component of the communication fabric, enabling real-time decision making, self-optimisation, and intent-driven orchestration.

The Open Radio Access Network (O-RAN) architecture enables intelligent network control through RAN Intelligent Controllers (RICs) that support components such as xApps and rApps, which are modular applications that can dynamically manage radio resources. O-RAN employs a disaggregated architecture that divides traditional gNodeB functionality across three open units: the Radio Unit (RU), Distributed Unit (DU), and Centralised Unit (CU). This allows mobile network operators (MNOs) to optimise performance, leverage cloud elasticity, and meet the diverse demands of different applications and services within the O-RAN framework \cite{qazzaz2025oran}. Individual components can be invoked manually by the MNO or, more implicitly, through Agentic AI frameworks to meet performance goals~\cite{wu2023autogen, Wu2025LLMDrivenAgentic}. However, integrating LLM-based models in O-RAN environments presents unique challenges due to task specific operational timescales and control loop requirements. These include ensuring trustworthy decision-making under strict latency and safety constraints, dealing with limited compute and memory resources at the edge, and managing the complexity of large-scale agent coordination.

\begin{figure}
    \centering
    \includegraphics[width=0.95\linewidth]{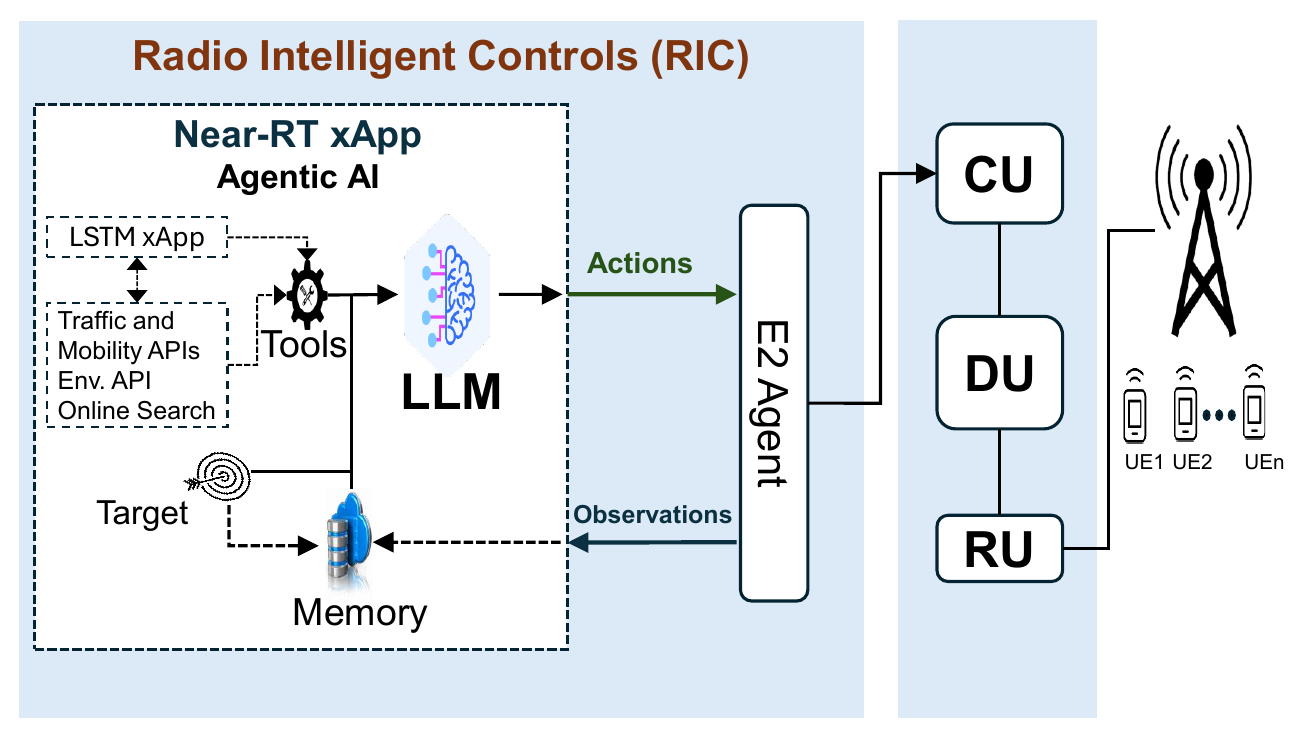}
    \caption{An O-RAN architecture featuring an xApp powered by an agent-based model.}
    \label{fig:O-RAN architecture featuring an xApp LLM}
\end{figure}

To address these challenges, we present an Agentic AI framework within edge-based RIC for network optimisation, as shown in Fig. \ref{fig:O-RAN architecture featuring an xApp LLM}. We present a novel Agentic AI framework that incorporates two key innovations: (i) a persona-based agent model and (ii) embedded search tools along with Long Short-Term Memory (LSTM) xApp for predictive traffic analytics. The persona-based architecture allows agents to simulate specialised roles, such as outage mitigator or energy optimiser, drawing from recent advances in contextual agent design for reasoning, dialogue, and control~\cite{Samuel2024Personagym, li2023chatharuhi, kong2023better}. Meanwhile, the integrated LSTM model continuously forecasts network load and signal quality trends over future time horizons. LSTM integration within the RIC enables proactive outage prevention, fine-grained power control, and early anomaly detection for near real-time edge operations.

Furthermore, a tool-aware reward design aligns the agent's decisions with system-level goals such as maintaining Signal-to-Interference-plus-Noise Ratio (SINR) thresholds, reducing outage rates, and optimising energy efficiency. The agent leverages O-RAN-compliant tools to implement decisions safely, adapting in real time to changing network and user contexts.

The main contributions of this work are as follows:

\begin{itemize}
    \item \textbf{RIC-Based Agentic AI Integration}: A lightweight, persona-driven agent model deployed within the near-RT RIC, enabling real-time, role-based decision-making for wireless optimisation at the edge.
    
    \item \textbf{Embedded Predictive LSTM Model}: A proactive traffic prediction module that forecasts SINR and traffic trends, enhancing the agent's planning and anomaly prevention capabilities.

    \item \textbf{Content-Aware Tooling \& Reward Design}: A novel agent architecture supported by content-aware tools and real-time information updates, implementing a reward mechanism that balances power consumption with QoS requirements and energy-saving objectives.

    \item \textbf{Autonomous Network Management}: The framework maps operator intents and network state into low-level configurations, enabling autonomous adaptation under dynamic operational conditions.
\end{itemize}

\section{Related work}
Contemporary wireless networks leverage diverse AI technologies for network optimisation and management. LLMs, including GPT-4, LLaMA, and domain-specific BERT variants, are being deployed by industry leaders such as NVIDIA, Ericsson, SoftBank, AT$\&$T, Nokia, and Vodafone for automated network operations \cite{SoftbankRedhat2025Power, Vasishta2024NvidiaAerial, Ericsson2025GenAI}. Autonomous AI agents capable of independent decision-making are being operationalised for complex network orchestration tasks \cite{Nvidia2025ARCCompact, Wu2025JointAdmission}.
Within the RAN Intelligent Controller (RIC) framework, reinforcement learning agents and hierarchical planners orchestrate xApps and rApps, forming the backbone of intelligent RAN operations \cite{Brik2024ExplainableAI, salama2025fedora}. Advanced agentic frameworks supporting multi-step planning and complex reasoning are under active investigation in both academic and industrial research contexts \cite{Samuel2024Personagym, Mao2025ALYMPICS}.
LLMs are trained or fine-tuned on telecom-relevant datasets, including standards documentation and troubleshooting logs~\cite{maatouk2023teleqna}.  
The work by Dev et al.~\cite{dev2025advanced} presents a forward-looking framework for integrating Agentic AI into 5G architectures, offering benefits such as simplified network operations through user/control plane separation, dynamic service orchestration via serverless computing, and enhanced adaptability using autonomous cognitive agents and neural radio protocol stacks. It highlights substantial gains in energy efficiency, latency reduction, and robustness, particularly in V2X scenarios.

LLMs exhibit strong coding and configuration capabilities. They enhance code quality and productivity in network software development~\cite{du2023power}, support intent-to-policy translation~\cite{dzeparoska2023llm}, and aid in generating and verifying network configurations~\cite{mondal2023llms}. Modular prompting strategies further improve LLM performance in complex telecom tasks~\cite{xiang2023toward}. Emerging strategies such as collaborative end-edge-cloud computing~\cite{zou2023wireless} and dynamic frameworks such as EdgeFM~\cite{yang2023edgefm} offer adaptive solutions, balancing performance, latency, and efficiency across heterogeneous network environments.

In prediction and classification, LLMs contribute to tasks such as traffic forecasting, signal processing, and security threat detection. Pre-trained models like BERT and GPT have been adapted to classify encrypted traffic and identify network anomalies~\cite{lin2022bert}.

LLMs also drive advances in telecom optimisation. They streamline reinforcement learning by automating reward function design~\cite{kwon2023reward}, and aid in solving convex and black-box optimisation problems~\cite{chen2024diagnosing}, showing promising results in resource allocation and system tuning.

\begin{figure*}
    \centering
    \includegraphics[scale=.53]{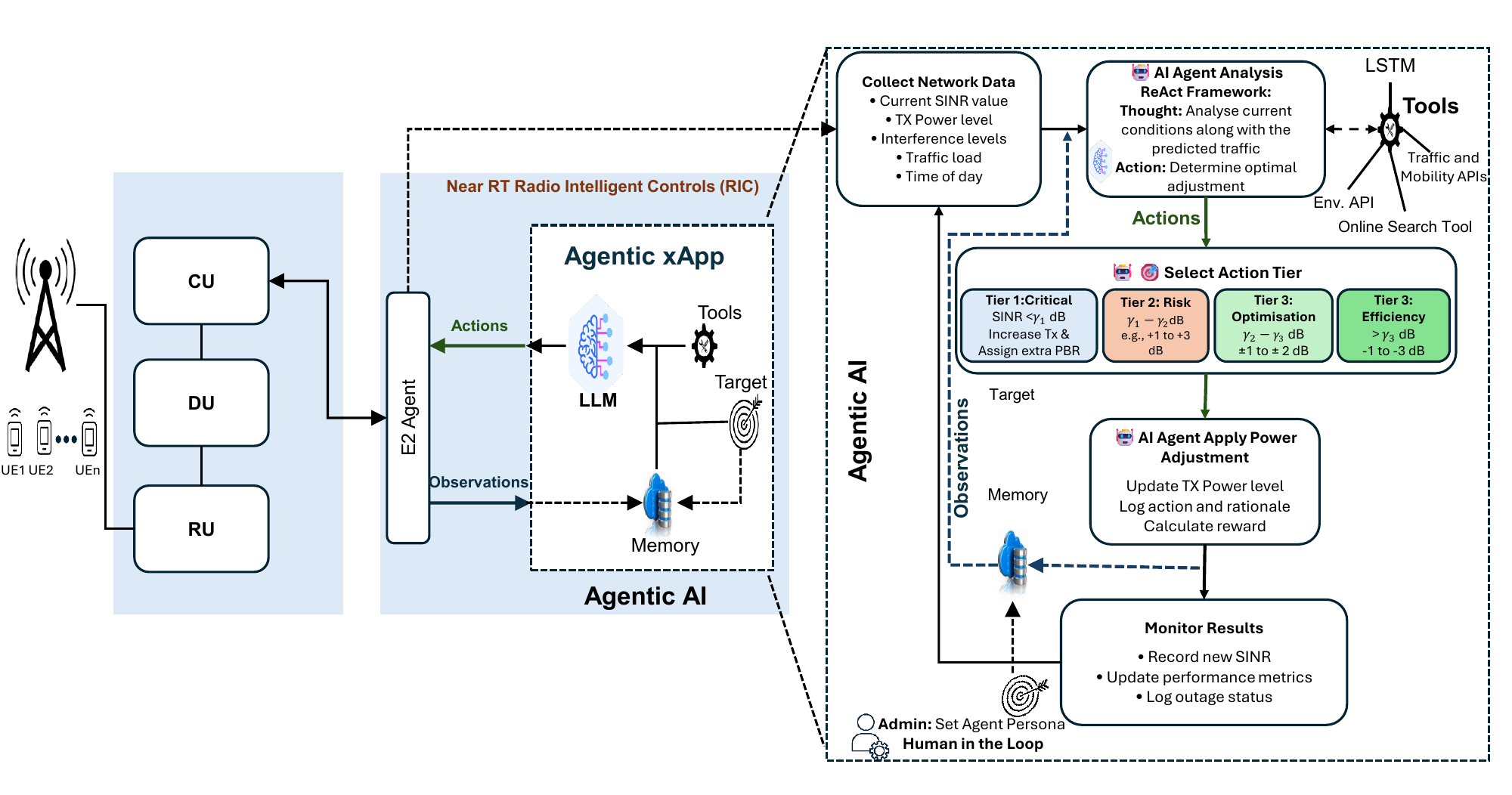}
    \caption{System Model Layout - AI Agent Framework for AI-RAN Network Optimisation.}
    \label{fig:system model}
\end{figure*}

Deployment remains a key challenge due to the high computational demands of LLMs. While cloud-based models offer scalability, they often suffer from latency and bandwidth limitations~\cite{shen2024large}. In contrast, edge and on-device deployments improve responsiveness but are constrained by limited processing and memory resources~\cite{ding2023parameter,kan2024mobile}. Efficient use of RIC capabilities can assist in overcoming this gap, enabling the deployment of reliable and robust GenAI-driven optimisation systems within resource-constrained network environments.

\section{System model}

Wireless network optimisation faces the ongoing challenge of maintaining optimal Quality of Service (QoS) under constantly changing conditions, including fluctuating traffic loads, interference patterns, and channel fading. Traditional approaches using predefined algorithms struggle to adapt to predictable yet complex scenarios that significantly impact network performance, such as national festivals where large crowds concentrate in limited areas, causing network overload. Our framework addresses these limitations by employing an Agentic AI framework for autonomous decision-making that monitors and controls networks in real-time. Our novel framework, illustrated in Fig.~\ref{fig:system model}, leverages a suite of tools to maintain real-time awareness of network conditions and proactively collaborate to predict and mitigate potential outages before they impact user experience. Our system comprises two integrated components operating in a closed-loop fashion:

\textbf{Component 1: Wireless Network Simulator} - Generates realistic 5G network data including KPIs, TX power levels, interference measurements, traffic loads, and temporal patterns. 

\textbf{Component 2: ReAct AI Agent} - Processes network state information and autonomously determines optimal power adjustments using LLM models and tools with a structured reasoning framework.
\subsection{Realistic Network Modelling}

The model is designed to replicate a realistic urban environment, incorporating comprehensive 5G characteristics:

\textbf{Traffic Dynamics}: Gaussian-distributed peaks during morning (7-9 AM, centred at 8:00) and evening (5-8 PM, centred at 18:30) with base load variations, augmented by special event modelling for crowd gatherings during festivals, concerts, and sporting events (up to 5× normal capacity demand).

\textbf{Multi-Source Interference}: Co-channel interference from distant cells (-110 dBm base), traffic-dependent variations (up to +10 dB during peaks), industrial interference (+3 dB during business hours), crowd-density interference (additional +2 to +8 dB in high-density areas), and random Gaussian variations ($\sigma = 2$ dB).

\textbf{Channel Effects}: Log-normal shadow fading (4 dB standard deviation) with exponential correlation ($\alpha = 0.98$), Rayleigh fast fading adjusted for mobility patterns, and comprehensive weather-induced variations including seasonal atmospheric effects, precipitation-based attenuation (up to 15 dB/km for heavy rain at mmWave frequencies), humidity-dependent absorption, and temperature-induced atmospheric ducting effects.

\textbf{mmWave-Specific Challenges}: Atmospheric oxygen absorption at 60 GHz band, human body blockage effects (10-20 dB attenuation) during crowd gatherings, foliage seasonal variations (2-8 dB additional loss), and enhanced weather sensitivity requiring adaptive beamforming and beam switching strategies.

\textbf{Event-Driven Scenarios}: Dynamic modelling of predictable network stress events, including national holidays, local festivals, stadium events, and emergency situations, where population density can increase by 200-800\% in localised areas, creating temporary hotspots requiring immediate resource allocation adjustments.

\subsection{AI Agent Tool Integration}
Our framework equips agents with diverse real-time data sources and control mechanisms:

\textbf{Event Intelligence}: Social media APIs (Twitter, Facebook Events) and event platforms (Ticketmaster, Eventbrite) enable proactive detection of crowd-gathering events, while traffic APIs (Google Maps, public transit) track population movement patterns.

\textbf{Environmental Intelligence}: Weather APIs (OpenWeatherMap, AccuWeather) provide precipitation, humidity, and atmospheric conditions critical for mmWave propagation modelling, while atmospheric monitoring services detect ducting conditions and seasonal foliage variations.

\textbf{Network Control Tools}: Software-defined networking APIs enable dynamic resource allocation, beamforming control systems manage antenna arrays, and load balancing APIs optimise traffic distribution across frequency bands.

The integration of external APIs in our framework raises potential reliability and security concerns. To mitigate these risks, we implement multi-source data validation by gathering information from different reliable sources and applying a correlation analysis agent to cross-verify data consistency across sources and identify discrepancies before passing it to our LSTM traffic prediction agent.

\subsection{AI Agent Decision Framework}

The ReAct (Reasoning and Acting) framework follows a systematic approach:

\textbf{Step 1 - Data Collection}: Agent receives network metrics and other related updated information via agent tools (SINR, power, interference, traffic, time).

\textbf{Step 2 - Intelligent Analysis}: Using ReAct prompting structure:
\begin{itemize}
\item \textit{Thought}: Analyse current network conditions and identify optimisation needs
\item \textit{Action}: Determine specific power adjustment strategy
\end{itemize}

\textbf{Step 3 - Tiered Decision Making}: Four-tier hierarchy, based on SINR ratio, ensures appropriate response:
\begin{itemize}
\item \textit{Tier 1 Critical} (SINR $< \gamma_1$ dB): Apply emergency power boost (increase transmission power by up to 3 dB) and allocate additional Physical Resource Blocks (PRBs) to maintain link stability and prevent imminent outage.

\item \textit{Tier 2 Risk} ($\gamma_1 \leq$ SINR $< \gamma_2$ dB): Initiate preventive adjustments (+1 to +3 dB) to avoid potential degradation in link quality.
\item \textit{Tier 3 Optimization} ($\gamma_2 \leq$ SINR $< \gamma_3$ dB): Perform fine-grained tuning ($\pm$1 to $\pm$2 dB) to balance performance with energy consumption.
\item \textit{Tier 4 Efficiency} (SINR $\geq \gamma_3$ dB): Reduce transmission power (-1 to -3 dB) to improve energy efficiency while maintaining acceptable QoS levels.
\end{itemize}
where $\gamma_1$ denotes the critical network threshold, set to 15 dB in this work to satisfy high QoS requirements. The remaining thresholds are defined as follows: $\gamma_2$ = 18 dB, $\gamma_3$ = 20 dB, and $\gamma_4$ = 25 dB.

\textbf{Step 4 - Action Implementation}: Apply power adjustment and monitor results.

\textbf{Step 5 - Performance Feedback}: Calculate the reward and update the system metrics and model memory.

The reward function incentivises SINR improvement while penalising excessive energy consumption:
\begin{equation}
R_{total} = 10 \cdot \Delta_{SINR} + R_{threshold} + R_{action} - 2 \cdot \Delta_{power}
\end{equation}

where $\Delta_{SINR}$ represents SINR improvement, $R_{threshold}$ provides threshold compliance rewards, $R_{action}$ encourages appropriate intervention, and $\Delta_{power}$ penalizes unnecessary power increases.

\textbf{System Assumptions}: 
The proposed system operates under the assumption that dynamic allocation of available PRBs and adaptive transmission power control are sufficient to maintain link reliability and service continuity, even under degraded network conditions. 

\subsection{LSTM xApp Integration into the Agentic AI Framework}

Our framework incorporates a novel multi-tool approach by embedding a trained Long Short-Term Memory (LSTM) xApp as a core predictive component within the Near Real-Time RIC. This integration advances predictive network management by combining LSTM's temporal modelling strengths with the contextual reasoning of LLM-driven agents. The LSTM xApp functions as a specialised forecasting tool within a collaborative agentic framework. It enhances traffic prediction accuracy by working alongside other data-driven tools, enabling real-time, context-aware insights. Unlike conventional LSTM models operating in isolation, our design fuses LSTM predictions with agent-derived contextual data for more informed decision-making. The integration operates through a three-stage collaborative pipeline:

\textbf{Stage 1 – Contextual Data Aggregation:} Agent tools continuously collect dynamic contextual signals, such as event schedules, location data, weather forecasts, and social media activity, to enrich the input features beyond basic historical metrics.

\textbf{Stage 2 – Predictive Modelling via LSTM:} The LSTM xApp processes both historical traffic data and real-time contextual features, producing probabilistic traffic forecasts with associated confidence intervals across short- and medium-term horizons.

\textbf{Stage 3 – Agent-Driven Interpretation:} LLM-based agents interpret the LSTM outputs in relation to current network conditions, optimising resource allocation strategies based on capacity, energy constraints, and service-level priorities.

The LSTM model employs a two-layer LSTM architecture (128, 64 units) followed by dense layers (32, 16 units). Training parameters include 500 epochs, batch size 32, learning rate 0.001, dropout 0.2, early stopping 10, and Adam optimiser.

\subsection{Key Performance Indicators}

The network metrics are based on the network performance and outage rate, where the model takes action in near-RT RIC (i.e., Agentic AI xApp) to satisfy the users' requirement, and these metrics are calculated as follows: 

\textbf{Outage Rate}: Percentage of time SINR falls below the threshold (i.e., $\gamma_1$ dB):
\begin{equation}
\text{Outage Rate} = \frac{\text{SINR} < \gamma_1\text{ dB samples}}{\text{Total samples}} \times 100\%.
\end{equation}

\textbf{Action Rate}: Percentage of significant power adjustments:
\begin{equation}
\text{Action Rate} = \frac{\text{\#Agent actions adjustments} }{\text{Total time steps}} \times 100\%.
\end{equation}

Furthermore, the model features novel persona-based AI agents with specialised roles in which strategic coordinators ensure long-term stability through scalable detection, tactical coordinators enable rapid action response, and balance the trade-off between performance and power consumption. Each persona agent employs tailored decision-making strategies and reward mechanisms, allowing the framework to autonomously adapt its anomaly detection and outage prevention based on operational context and service provider priorities.

\section{Results and Performance Analysis}

To evaluate the proposed model, we simulate a realistic urban 5G environment incorporating high-fidelity parameters such as 43 dBm base power, 120 dB path loss, and a thermal noise floor of -104 dBm. The scenario captures complex real-world dynamics, including time-dependent traffic peaks, multi-source interference, diverse channel effects, and mmWave-specific propagation challenges. Event-driven scenarios such as festivals and emergencies are also modelled, introducing significant, localised network stress. This comprehensive setup ensures robust testing of the model’s performance under highly variable and demanding conditions.

Our system successfully identifies and responds to critical periods of abnormal high traffic peaks, as shown in Fig. \ref{fig:LSTM prediction}, demonstrating predictive and proactive optimisation capabilities through the proposed agent tools (e.g., online search API tool). These tools collaborate with the xApp-integrated LSTM model by providing updates in real-time related information (e.g., event time and location, estimated event traffic capacity) to deliver high-accuracy, near real-time traffic predictions, enabling the agent to apply appropriate network configurations that maintain stable performance and meet user requirements. The LSTM model's predictive accuracy makes it suitable for real-time network capacity planning and resource allocation strategies. The 98\% prediction accuracy, illustrated in Fig. \ref{fig:LSTM prediction}, enables network operators to proactively adjust bandwidth allocation, optimise load balancing, and implement dynamic resource scaling based on anticipated traffic demands, particularly in critical situations such as event times.

\subsection{Baseline Comparison and Performance Metrics}

Three distinct approaches are evaluated to establish comprehensive performance benchmarks, as shown in Fig. \ref{fig: AI Agent Results LLM}. Traditional fixed-power networks rely on static configurations based on historical averages without adaptive capabilities. Standard LLM-based reactive models make decisions based solely on current network measurements with limited contextual awareness. Our proposed Edge AI Gen model employs proactive decision-making with multi-source data integration and predictive capabilities. Performance evaluation focuses on three critical metrics: outage prevention effectiveness, SINR stability maintenance, and system responsiveness.

\begin{figure}
    \centering
    \includegraphics[width=0.95\linewidth]{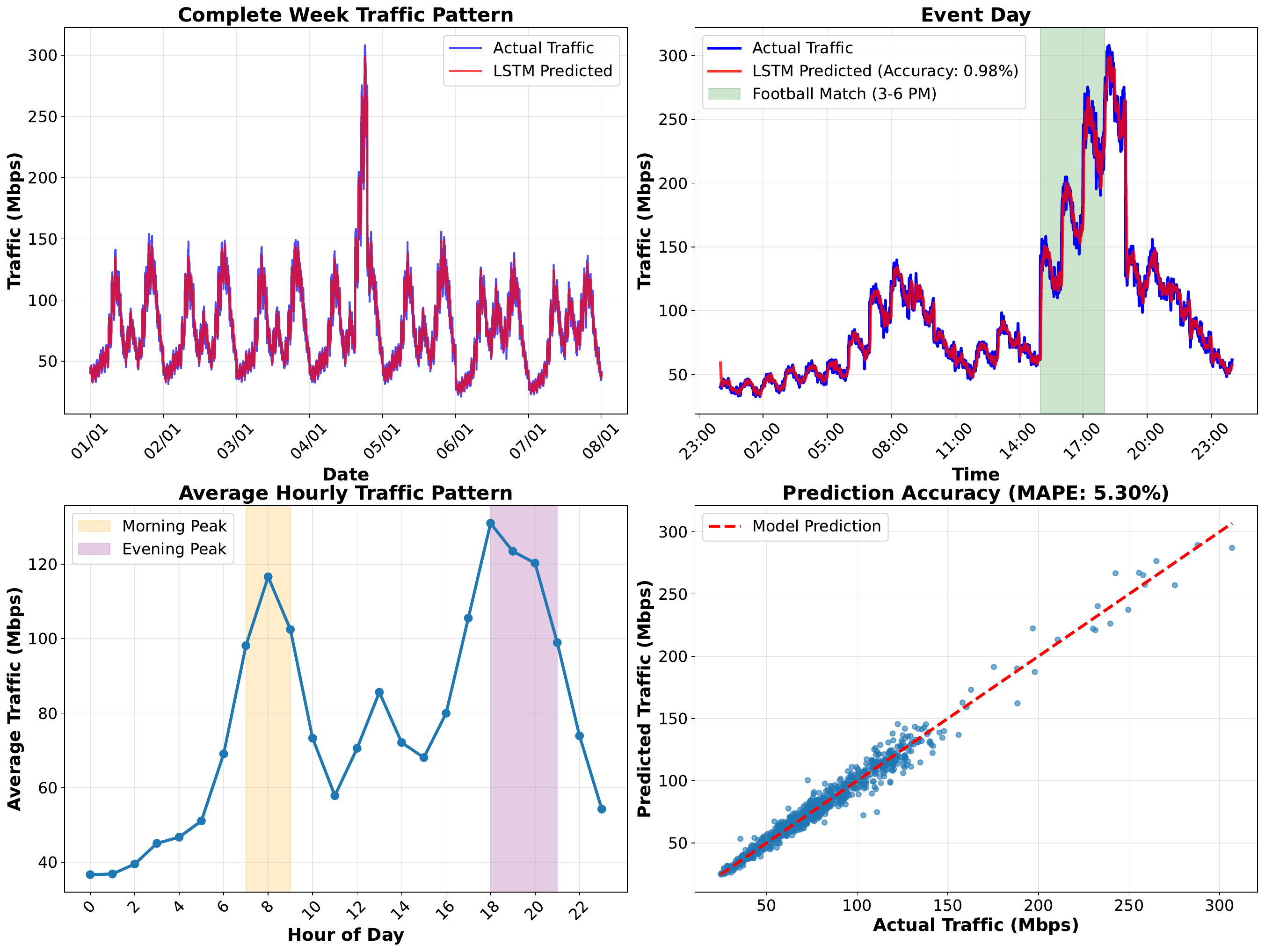}
    \caption{LSTM model prediction results in wireless network traffic forecasting, effectively capturing traffic load patterns over a 7-day prediction horizon.}
    \label{fig:LSTM prediction}
\end{figure}

As illustrated in Fig. \ref{fig:LSTM prediction} and \ref{fig: AI Agent Results LLM}, the most compelling validation of our approach emerges during the peak hours and special event scenarios, where large gatherings triggered extreme demand spikes and elevated interference levels. Under these challenging conditions, our model is evaluated over a high-traffic day, using minute time steps, characterised by abnormal load spikes due to overlapping peak hours and concurrent event activity. It achieves a 100\% success rate in preventing network outages while maintaining stable performance across all user connections. This exceptional performance is attributed to its context-aware operation, where proactive persona agents leverage specialised tools to estimate demand in advance and take timely, appropriate actions. In contrast, baseline methods under the same conditions experience high outage rates: 8.4\% for traditional fixed-power networks and 3.3\% for standard LLM agent-based models. 

Therefore, the key enabler for this performance breakthrough lies in the model's proactive and context-aware capabilities. Unlike reactive systems that respond only after degradation occurs, our Edge AI Gen model continuously integrates real-time data from multiple sources, including social media feeds, municipal event calendars, and weather monitoring systems. This multi-modal data fusion enables the model to anticipate traffic surges and prepare network infrastructure proactively, adjusting power levels before events begin rather than responding to congestion after it manifests.

Quantitative analysis of SINR performance demonstrates substantial improvements in network quality metrics. During special event day, when interference levels typically increase due to elevated device density, our model achieves a significant reduction in SINR variance compared to baseline approaches. This translates to measurably improved user experience, with 0\%  service disruption reports and more stable connections in high-density environments, as shown in Fig. \ref{fig: AI Agent Results LLM} and \ref{fig:Radar Chart}.

\subsection{Edge Deployment Advantages and Real-time Performance}

\begin{figure}
    \centering
    \includegraphics[width=0.98\linewidth, height=1.45\linewidth]{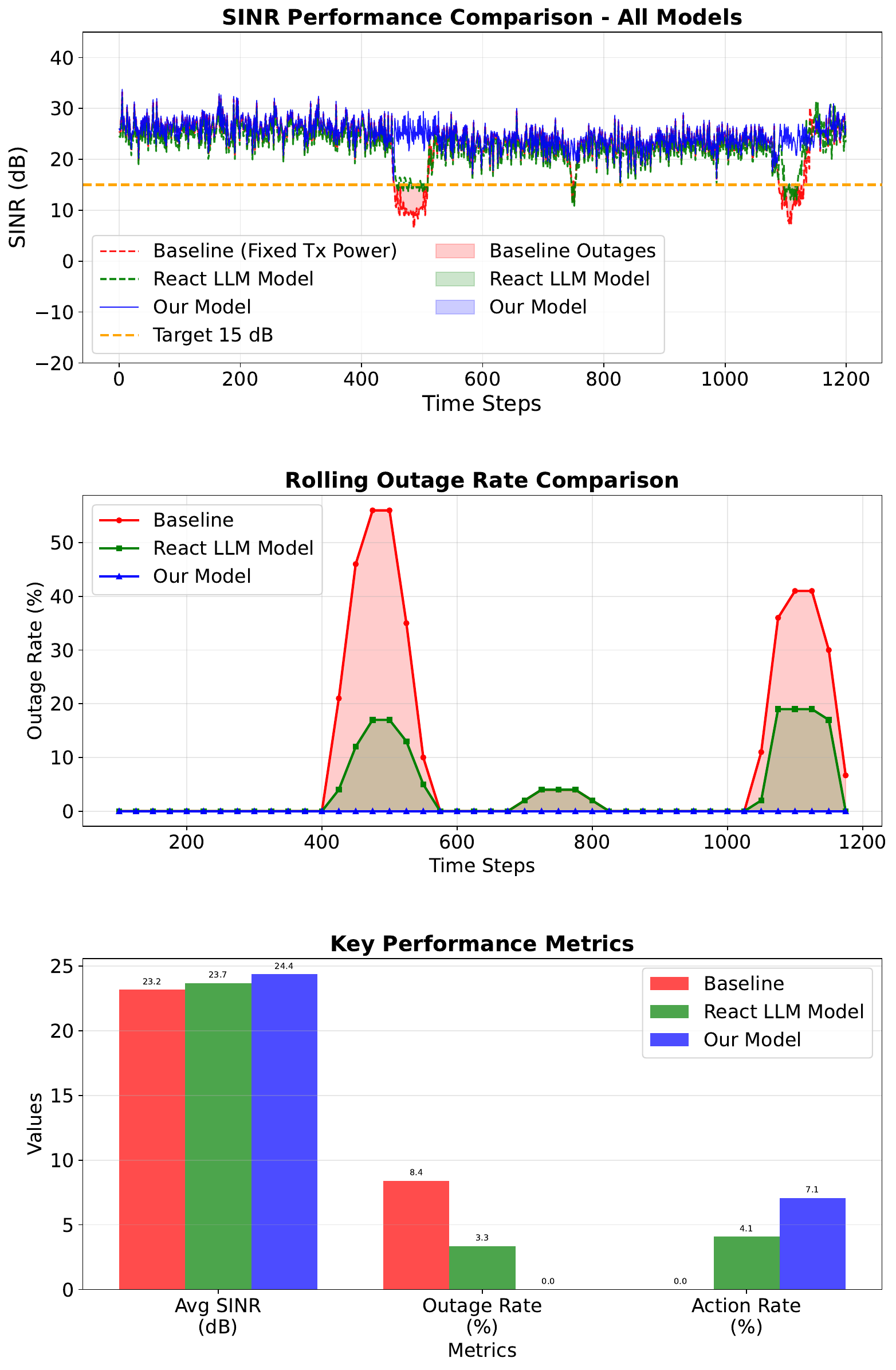}
    \caption{Performance comparison of models for anomaly detection and outage prevention in wireless networks under high traffic demand.}
    \label{fig: AI Agent Results LLM}
\end{figure}

\begin{figure}
        \centering
        \includegraphics[width=0.98\linewidth, height=0.65\linewidth]{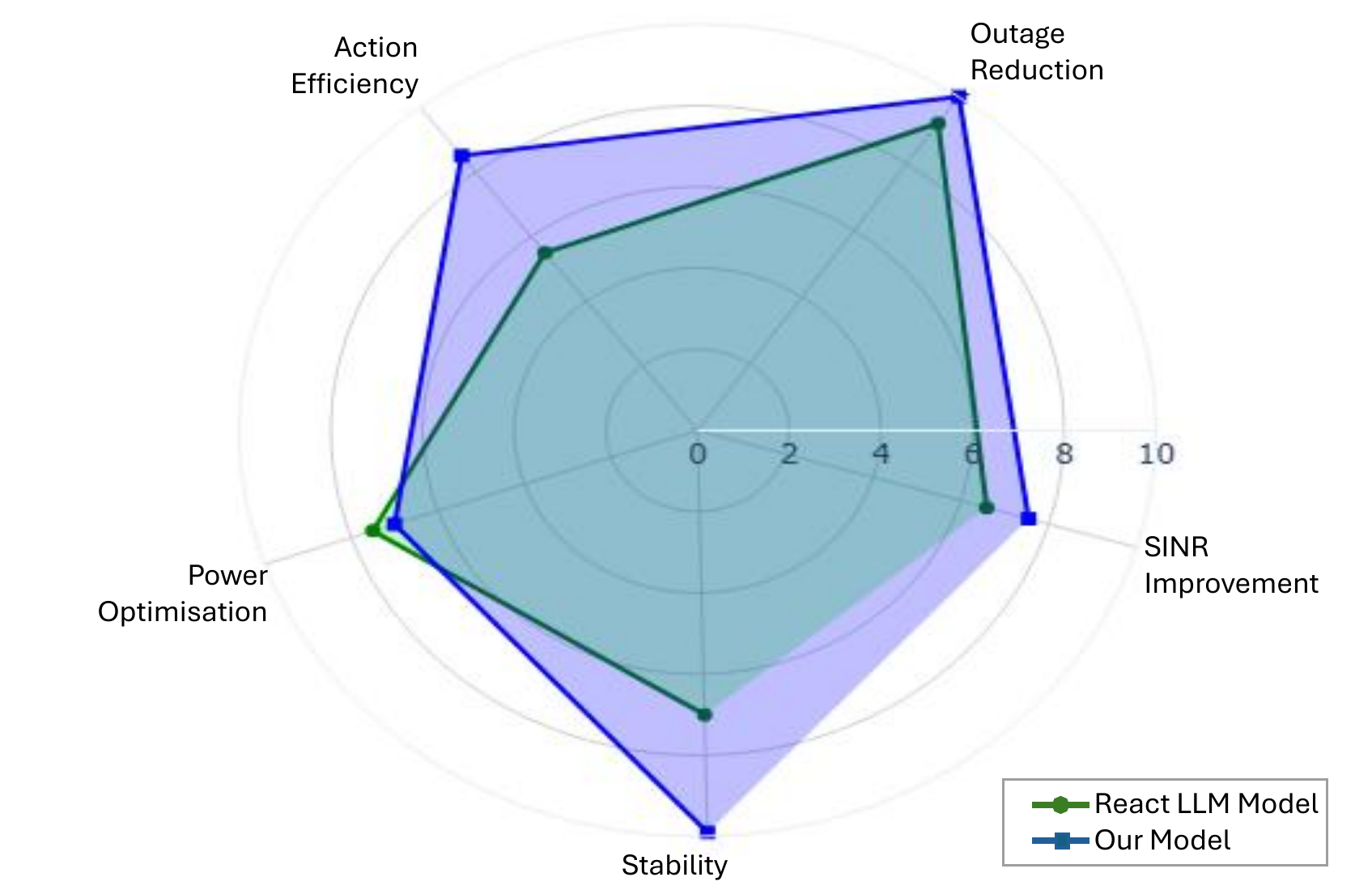}
        \caption{Radar chart comparing the performance of our proposed model against the baseline LLM-based model across key operational metrics.}
        \label{fig:Radar Chart}
\end{figure}

Performance improvements demonstrated robust validation across all evaluation metrics, as shown in Fig. \ref{fig: AI Agent Results LLM} and \ref{fig:Radar Chart}. Model validation is conducted using minute-level granular data generated over a 2-month training period, a minute time step (total 86,400 time steps), with performance evaluated on 15 randomly selected event days representing diverse anomaly scenarios and critical operational periods. Our Edge GenAI model achieved superior traffic load prediction with 98\% accuracy.

Critical performance evaluation during high-stress event scenarios (i.e., high traffic demand) demonstrates 100\% outage prevention success rate in compared to 91.6\% (traditional) and 96.7\% (standard LLM-based baseline). The model maintained a coefficient of variation below 0.15 across all metrics during the 15 validation event days, showing operational consistency.

Therefore, integration with RIC architecture provides critical advantages for real-time network management. Edge deployment ensures optimisation decisions are implemented immediately without centralised processing delays, which proves essential during rapidly evolving network conditions. The edge deployment also enables access to current network state information with minimal latency, supporting more accurate and timely intervention decisions. This architectural advantage, combined with generative AI capabilities and agentic tool utilisation, creates a fundamentally more intelligent network management paradigm that reasons about complex scenarios, predicts future network states, and implements preventive measures before user impact occurs.


While our results demonstrate significant improvements with strong statistical support, some limitations warrant consideration for future development. Scalability across diverse network topologies and operator environments needs further validation through large-scale deployments.

\section{Conclusion}

This paper introduces a comprehensive Edge AI framework for autonomous network optimisation in O-RAN environments, addressing critical challenges in 5G and beyond through intelligent, persona-based agents deployed within the RIC infrastructure. Our approach advances proactive network management by moving beyond traditional reactive strategies, enabling predictive, context-aware decision-making. The proposed system demonstrates unprecedented levels of situational awareness and optimisation capability. It achieves a 100\% success rate in preventing network outages during high-demand scenarios such as special events, where baseline methods suffer significant service degradation, while also improving response times and adaptability. These results validate the effectiveness of our framework across multiple performance dimensions. The integration of advanced AI capabilities with real-world network management requirements highlights the potential for significant gains in operational resilience and efficiency. This work contributes to the broader vision of fully autonomous, self-optimising networks that dynamically adapt to changing conditions without human intervention. As the industry evolves toward intelligent, scalable, and reliable network services, Edge AI frameworks like ours will be foundational to realising the next generation of wireless communication.

\section*{Acknowledgment}
This research was funded by EPSRC CHEDDAR (EP/X040518/1), UKRI Grant EP/X039161/1, and MSCA Horizon EU Grant 101086218. Additionally, this research was supported by the UKRI Funding Service under Award UKRI851: Strategic Decision-Making and Cooperation among AI Agents: Exploring Safety and Governance in Telecom.

\bibliographystyle{unsrt}
\bibliography{References}

\newpage


\vfill

\end{document}